\documentclass[prb,preprint,superscriptaddress,floatfix]{revtex4-1}
\usepackage{amsmath}  % needed for \tfrac, \bmatrix, etc.
\usepackage{amsfonts} % needed for bold Greek, Fraktur, and blackboard bold
\usepackage{graphicx} % needed for figures
\usepackage{mathptmx, bm}
\usepackage{float}

\begin{document}

\title{An elementary derivation of first and last return times of 1D random walks}

\author{Sarah Kostinski}
\email{skostinski@physics.harvard.edu}
\affiliation{Department of Physics, Harvard University, 17 Oxford St, Cambridge, MA 02138}

\author{Ariel Amir}
\email{arielamir@seas.harvard.edu}
\affiliation{School of Engineering and Applied Sciences, Harvard University, 29 Oxford St, Cambridge, MA 02138}

%\date{\today}

\begin{abstract}
Random walks, and in particular, their first passage times, are ubiquitous in nature. Using direct enumeration of paths, we find the first return time distribution of a 1D random walker, which is a heavy-tailed distribution with infinite mean. Using the same method we find the last return time distribution, which follows the arcsine law. Both results have a broad range of applications in physics and other disciplines. The derivation presented here is readily accessible to physics undergraduates, and provides an elementary introduction into random walks and their intriguing properties.

\end{abstract}

\maketitle

Thermal and statistical physics texts often begin with a discussion of random walks\cite{reif, kittel, krapivsky} and their associated applications such as Brownian motion,\cite{brownian} polymer physics,\cite{doi} and laser cooling of atoms.\cite{coldatoms}  The first passage time distribution $F(r,t)$, i.e. the distribution of times $t$ at which a target $r$ is first reached, stands out as central to many natural phenomena such as quenching of fluorescent molecules,\cite{cichos} molecular rupture (times over which molecules dissociate in, e.g., ligand-receptor complexes), and target site searches (e.g. transcription factors finding corresponding binding sites along DNA),\cite{chou} as well as additional problems in biology.\cite{zilman} In the context of finance, an optimal trading strategy might be to sell an asset when it first reaches a threshold value.\cite{finance, bouchaud}

Insofar as undergraduate texts give an impression that the bell-shaped curve rules the world, first passage time distributions supply a nice counterexample by their heavy-tailed behavior, e.g., $\sim t^{-3/2}$ in 1D. Here we supply an elementary derivation of this result by examining first returns on an infinite 1D lattice. While general results\cite{feller, redner} are available for first return distributions on infinite $d$-dimensional lattices, derivations rely on generating functions or Laplace transforms which may be unfamiliar to undergraduates. In contrast, the approach below yields the power law after just a few lines of mathematics by entirely elementary means. It can thus serve as a friendly primer to random walks, first passage, recurrence, and heavy-tailed distributions.

\begin{figure}[hb]
\centering
\includegraphics[width=0.45\textwidth]{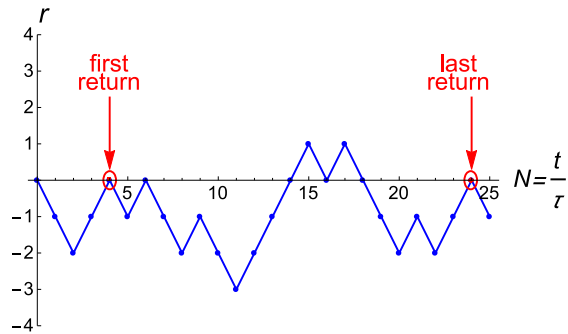}
\caption{\textit{First and last returns}: The above plot is an example 1D random walk of $N=25$ steps beginning at the origin $r=0$ at time $t=0$. The ordinate and abscissa are the distance $r$ traveled and time $t/\tau$ elapsed (number of steps taken), respectively. The first and last returns to the origin are marked in red. In this case, the first return time is $N_{\text{first}} = t/\tau = 4$, while the last return time is $N_{\text{last}} = t/\tau = 24$.}
\label{exampleRW}
\end{figure}

A 1D random walk is a succession of $N$ steps to the right or left with respective probabilities $p$ and $q=1-p$, occurring at every time interval $\Delta t = \tau$ (hence $N=t/\tau$). We focus on the case of a symmetric walk ($p=q=1/2$) but the same formalism may be applied to biased walks ($p\neq q$). All walks considered here begin at the origin ($r=0$) at time $t=0$ and have steps of identical length $1$. The first return time is the time at which the walk first reaches the origin; similarly, the last return time is the time at which the origin is last visited. See Fig.~\ref{exampleRW} for an example of a 1D random walk trajectory with its first and last returns marked.

Assuming spatial and temporal initial conditions $(r,t)=(0,0)$, the first return time distribution is $F(r=0,t)$, denoted as $F(t)$ hereafter. Its cumulative $\int_{0}^{t} F(t') dt'$ is the probability to return to the origin by time $t$. The complement is the survival probability $S(t)$, i.e. the probability to not return by time $t$:
\begin{equation}
\int_{0}^{t} F(t') dt' = 1-S(t) \; \longrightarrow \; F(t) = -\frac{\partial S(t)}{\partial t}.
\label{eq:deriv}
\end{equation}
$S(t)$ is found by enumerating all survival paths (those not returning to the origin) in the first $N$ steps. The probability of such a path occurring is given by the ballot theorem: In a ballot where candidates A and B have $a$ and $b$ total votes, respectively, the probability that A is always ahead of B throughout counting is $(a-b)/(a+b)$. To enumerate survival paths we use a proof of the ballot theorem, known as the cycle lemma.\cite{renault} The latter's cyclical representation of paths is ideal for counting their partial sums and eliminating those which return to the origin, as explained below.
\begin{figure}[h!]
\centering
\includegraphics[width=0.3\textwidth]{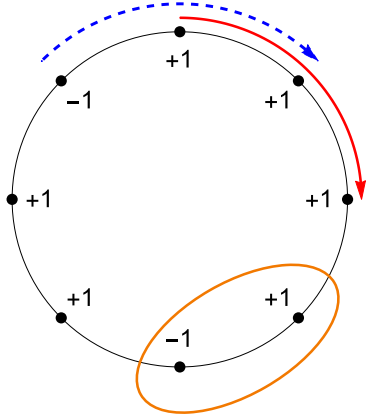}
\caption{\textit{Enumerating survival paths which remain to the right of the origin}: $N$ numbers are written on a circular track where each $+1$ and $-1$ denote a step to the right and left, respectively. The solid red and dashed blue arrows represent the start of two possible clockwise paths. The red (solid) path is a survival path, i.e. does not return to the origin, whereas the blue (dashed) path is not a survival path as its running sum along the path is not always positive. The orange oval highlights a \{+1,-1\} pair whose constituent numbers do not start survival paths. Furthermore, the pair's sum is zero and therefore it may be removed from the circular track without affecting any path's running sum.}
\label{circulartrack}
\end{figure}

Consider a circular track with $N$ numbers, a fraction of which are $+1$ and the rest are $-1$ (see Fig.~\ref{circulartrack}). The numbers $+1$ and $-1$ signify a step to the right and left, respectively. Such a configuration has $N$ possible clockwise paths along the circular track starting at each of the $N$ numbers. Consider first those survival paths which remain to the right of and never cross the origin: the sum of numbers along such paths is always positive. Any $+1$ followed by a $-1$ may be eliminated from the circular track as the two paths starting at either number are not in this class of right-of-the-origin survival paths; furthermore, their removal does not affect any other path's sum since a $\{+1, -1\}$ pair's net sum is zero. Repeating this procedure until no $-1$'s remain yields the number of $+1$'s in excess of $-1$'s, i.e. the number of valid paths. The probability of choosing a valid path from a given track is then
\begin{equation}
\frac{N_{+} - N_{-}}{N} = \frac{(N-N_{-}) - N_{-}}{N} = \frac{N-2N_{-}}{N}
\label{eq:probvalid}
\end{equation}
where $N_{+}$ and $N_{-}$ denote the number of $+1$'s and $-1$'s, respectively. The probability in Eq.~\ref{eq:probvalid} is non-negative since $N_{+} > N_{-}$ for a path to remain to the right of the origin. To obtain the total number of valid paths from all possible circular track configurations, Eq.~\ref{eq:probvalid} is multiplied by the number of possible $\{+1, -1\}$ arrangements $\binom{N}{N_{-}}$. The result is the ballot theorem: For a given $N_{-}$, the number of paths remaining to the right of the origin is $\frac{N-2N_{-}}{N} \binom{N}{N_{-}}$. Because $N_{+}$ must exceed $N_{-}$ for the walker to remain to the right of the origin, $N_{-}$ can range from 0 to $\left\lfloor \frac{N-1}{2} \right\rfloor$ where the latter floor function denotes the largest integer not greater than $\frac{N-1}{2}$. Summing over these values yields the number of paths which remain to the right of the origin:
\begin{equation}
\sum_{N_{-}=0}^{\left\lfloor \frac{N-1}{2} \right\rfloor}  \frac{N-2N_{-}}{N} \binom{N}{N_{-}} = \binom{N-1}{\lfloor N/2 \rfloor}.
\label{eq:Rsum}
\end{equation}
The summation in Eq.~\ref{eq:Rsum} is simplified by binomial identities $\binom{n}{x} = \binom{n-1}{x-1} + \binom{n-1}{x}$ and $x \binom{n}{x} = n \binom{n-1}{x-1}$ as follows:
\begin{equation}
\begin{split}
& \sum_{N_{-}=0}^{\lfloor \frac{N-1}{2} \rfloor} \binom{N}{N_{-}} - \frac{2}{N} \sum_{N_{-}=0}^{\lfloor \frac{N-1}{2} \rfloor} N_{-} \binom{N}{N_{-}} \\
= & \sum_{N_{-}=0}^{\lfloor \frac{N-1}{2} \rfloor} \left[ \binom{N-1}{N_{-}-1} + \binom{N-1}{N_{-}} \right] - \frac{2}{N} \sum_{N_{-}=0}^{\lfloor \frac{N-1}{2} \rfloor}  N \binom{N-1}{N_{-}-1}\\
= & \sum_{N_{-}=0}^{\lfloor \frac{N-1}{2} \rfloor} \binom{N-1}{N_{-}} - \sum_{N_{-}=0}^{\lfloor \frac{N-1}{2} \rfloor} \binom{N-1}{N_{-}-1} = \binom{N-1}{\lfloor \frac{N-1}{2} \rfloor}.
\end{split}
\label{eq:expandRsum}
\end{equation}
Furthermore, $\left\lfloor \frac{N-1}{2} \right\rfloor = \left \lceil \frac{N}{2} \right \rceil -1 \;$ which leads to the final result of Eq.~\ref{eq:Rsum}.
%Eq.~\ref{eq:expandRsum} is a difference of sums of adjacent entries in row $(N-1)$ of the left side of Pascal's triangle; one sum contains an extra term . All summation terms except one, corresponding to the upper summation limit $\lfloor \frac{N-1}{2} \rfloor$, cancel.
%\begin{equation}
%\binom{N-1}{\lfloor \frac{N-1}{2} \rfloor} = \binom{N-1}{\lceil N/2 \rceil -1} = \binom{N-1}{\lfloor N/2 \rfloor}.
%\end{equation}
By symmetry, the number of paths which remain to the left of the origin is the same as in Eq.~\ref{eq:Rsum}. Thus the number of survival paths is $2 \binom{N-1}{\lfloor N/2 \rfloor}$. The survival probability $S(N)$ follows simply; each path has probability $(1/2)^{N}$ and therefore
\begin{equation}
S(N) = 2^{-(N-1)} \binom{N-1}{\lfloor N/2 \rfloor}.
\label{eq:Sdiscrete}
\end{equation}
In the continuum limit where $N$ is large ($t \gg \tau$), Stirling's approximation $N! \sim \sqrt{2\pi N} (N/e)^{N}$ for Eq.~\ref{eq:Sdiscrete} gives $S(t)$:
\begin{equation}
S(N) = \sqrt{\frac{2}{\pi N}} \; \rightarrow \; S(t) = \sqrt{\frac{2 \tau}{\pi t}}.
\label{eq:Scont}
\end{equation}
The survival probability decays to zero for long times $t \rightarrow \infty$, implying that the walk will eventually return to the origin with probability 1. This is in accord with P\'olya's recurrence theorem\cite{polya, snell} that symmetric random walks return to the origin on infinite lattices of dimension $d\leq2$.

\begin{figure}[h!]
\centering
\includegraphics[width=0.45\textwidth]{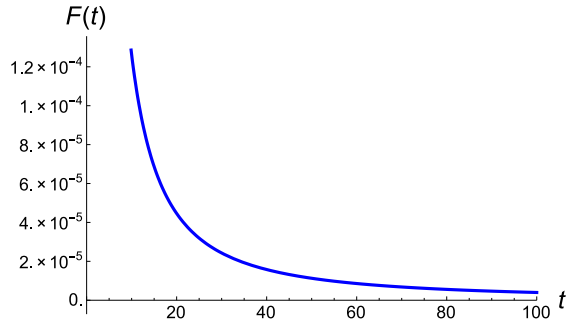}
\caption{\textit{Heavy-tailed distributions}: The first return time distribution $F(t) = \sqrt{\frac{\tau}{2 \pi t^{3}}} \sim t^{-3/2}$ is shown above, where $\tau=10^{-4} \ll t$. It is an example of a heavy-tailed distribution, which is typical for first passage time distributions. The average return time diverges; long return times in this distribution's heavy tail dominate the average.}
\label{heavytail}
\end{figure}
By Eq.~\ref{eq:deriv}, the first return distribution is
\begin{equation}
F(t) = -\frac{\partial}{\partial t} \left( \sqrt{\frac{2 \tau}{\pi t}} \right) = \sqrt{\frac{\tau}{2 \pi t^3}} \; \rightarrow \; F(t) \sim t^{-3/2} \, .
\label{eq:F}
\end{equation}
It follows that the distribution's first moment, the average return time, diverges:
\begin{equation}
\left< t_{return} \right> = \int_{0}^{\infty} t F(t) dt  \propto \int_{0}^{\infty} t \cdot t^{-3/2} dt = \infty.
\end{equation}
Diverging moments are the hallmark of heavy-tailed distributions; in this case, long return times dominate the average. Fig.~\ref{heavytail} shows the heavy tail distribution of $F(t) \sim t^{-3/2}$.
\begin{figure}[h!]
\centering
\includegraphics[width=0.45\textwidth]{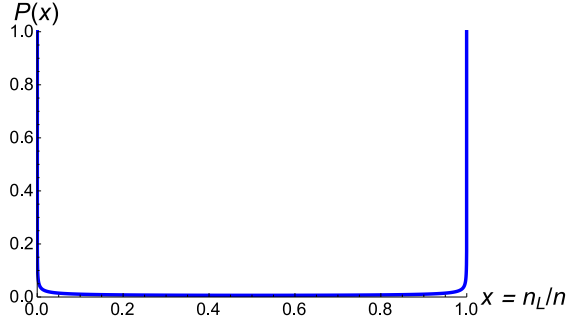}
\caption{\textit{Last return times}: The probability of returning to the origin for the last time at $x=n_{L}/n$ for $n=100$. Note its symmetric behavior about the minimum $x=1/2$. Last returns are much more likely to occur either very early or late in the walk.}
\label{lastreturn}
\end{figure}

\begin{figure}[ht]
\centering
\includegraphics[width=0.43\textwidth]{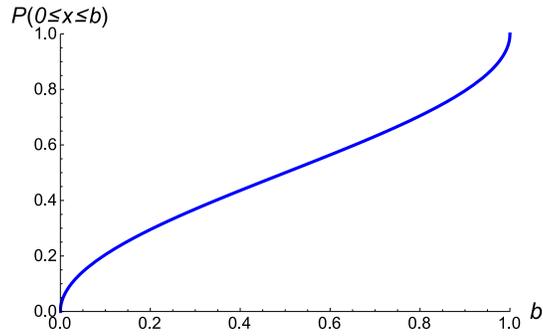}
\caption{\textit{The arcsine law}: The probability that the last return occurs within the first fraction $b$ of the full walk duration has the form $P(0 \leq x \leq b)=(2/\pi) \arcsin (\sqrt{b})$.}
\label{arcsine}
\end{figure}
Our derivation yields insight into last return times as well. The probability to return for the last time at step $2n_{L}$ (an even number of steps implied) is the product of the probabilities to be at the origin at step $2n_{L}$ and of surviving $2n-2n_{L}$ steps thereafter. The former is $2^{-2n_{L}} \binom{2n_{L}}{n_{L}}$ since there are $\binom{2n_{L}}{n_{L}}$ ways to take an equal number of steps right and left. For large $n_{L}$ the latter probability becomes $1/\sqrt{\pi n_{L}}$ by Stirling's approximation. Multiplying by survival probability $1/\sqrt{\pi (n-n_{L})}$ yields the probability that the last return occurs at step $2n_{L}$:
\begin{equation}
\frac{1}{\pi \sqrt{n_{L} (n-n_{L})}} = \frac{1}{\pi n \sqrt{x(1-x)}}
\label{eq:lastreturn}
\end{equation}
where $x = n_{L}/n$. Eq.~\ref{eq:lastreturn} is symmetric about its minimum $x=1/2$ with singular maxima occurring at $x=0$ and $x=1$ (see Fig.~\ref{lastreturn}). Integrating Eq.~\ref{eq:lastreturn} yields the arcsine law\cite{feller} $P(0 \leq x \leq b)=\frac{2}{\pi} \arcsin (\sqrt{b})$ as shown in Fig.~\ref{arcsine}. The arcsine law also describes the number of positive partial sums in a sequence of mutually independent random variables from probability distributions other than the binomial.\cite{erdos} While this rather counterintuitive result is seldom encountered in physics texts, the law has striking consequences likely to excite physics students. Feller\cite{feller} (see Vol. 1, Section III.4) describes it in the context of coin-tossing games where losing and winning sides map to the left and right of the origin, respectively, and equalization of the fortunes signifies a return to the origin:
\begin{quote}
The results are startling. According to widespread beliefs a so-called law of averages should ensure that in a long coin-tossing game each player will be on the winning side for about half the time, and that the lead will pass not infrequently from one player to the other. Imagine then a huge sample of records of ideal coin-tossing games, each consisting of exactly $2n$ trials. We pick one at random and observe the epoch of the last tie... \textit{With probability 1/2 no equalization occurred in the second half of the game, regardless of the length of the game.} Furthermore, the probabilities near the end points are \textit{greatest}... These results show that intuition leads to an erroneous picture of the probable effects of chance fluctuations.
\end{quote}
The last return distribution is tied to the time spent on either side of the origin, which also follows the arcsine law.\cite{feller} It is highly probable to remain on one side of the origin for nearly the entire walk, leading to long waiting times. Recent implications include hard-spheres gas particles colliding with the same neighbors for an extended period of time.\cite{fouxon} Other examples where the arcsine law is obeyed include the time of maximal displacement in 1D Brownian motion,\cite{comtet, levy} lead changes within competitive team sports games,\cite{NBA} and the probability distribution of longitudinal displacements of tracer particles in split flow.\cite{splitflow}

In summary, we have reported on an elementary derivation of first and last return times which also serves as an introduction to a variety of important and broadly applicable concepts such as recurrence, first passage, heavy-tailed distributions, and the arcsine law.

\begin{acknowledgments}
%\vspace{-2mm}
We thank Sidney Redner, Ori Hirschberg, and Michael P. Brenner for helpful comments. S.K. was supported by the U.S. Department of Defense through the NDSEG Program.
\end{acknowledgments}

\end{document}